# If Experts Converge on the Same Answer are they Less Creative than Beginners? Redefining Creativity in Terms of Adaptive Landscapes

Liane Gabora

University of British Columbia

June, 2011

Address for Correspondence:

L. Gabora <liane.gabora@ubc.ca>

Department of Psychology

University of British Columbia

Okanagan Campus

Arts Building, 3333 University Way

Kelowna BC, V1V 1V7

CANADA

## Abstract

The standard view that creativity entails both originality and appropriateness leads to the paradox that experts who converge on one optimal solution are rated as no more creative than beginners who give many original solutions. This paper asserts that there is no one-size-fits-all definition of creativity; creativity must be assessed relative to the constraints and affordances of the task. The flatter the adaptive landscape associated with the task, the greater the extent to which creativity is a function of originality only. For tasks with a single-peaked adaptive landscape, there is a tradeoff between originality and appropriateness. Only for tasks with rugged adaptive landscapes is creativity positively correlated with both originality and appropriateness. It is suggested that the adaptive landscapes associated with artistic and scientific pursuits are equally rugged, but for artistic pursuits their topologies reflect idiosyncratic experiences and emotions (the peaks and valleys are not aligned).

# If Experts Converge on the Same Answer are they Less Creative than Beginners? Redefining Creativity in Terms of Adaptive Landscapes

## 1. Introduction

The elusiveness of the construct of creativity makes it that much more important to obtain a satisfactory definition of it. Defining creativity presents difficulties; for example, not all creative works are useful, and not all are aesthetically pleasing, though both usefulness and aesthetic value capture, in some sense, what creativity is about. Nevertheless, psychologists have almost universally converged on the definition originally proposed by Guilford over sixty years ago (Feist, 2011; Moran, 2011). Guilford (1950) defined creativity in terms of two criteria: *originality* or novelty, and *appropriateness* or adaptiveness, i.e. relevance to the task at hand. *Surprise* is sometimes added as a third criterion (Boden, 2004). Some add *quality* as a separate criterion (Kaufman & Sternberg, 2007), while others use the term appropriateness in a way that encompasses quality. Creativity has also been defined as a complex (Albert & Runco, 1989) or syndrome (MacKinnon, 1975; Mumford & Gustafson, 1988), and some would insist that any definition of creativity include such cognitive and personality characteristics as problem sensitivity, flexibility, the ability to analyze, synthesize, evaluate, and reorganize information, engage in divergent thinking, or deal with complexity. However, it is the 'originality and appropriateness' definition that is encountered most often, and that appears to have become standard (e.g. Amabile, 1996; Campbell, 1960; Feldman, Csikszentmihalyi & Gardner, 1994; Runco, 2004; Sternberg, 1988).

While this definition provides a much-needed departure point for discussion about and measurement of creativity, it presents the following problem.[1] Consider a group of experts and a group of beginners each confronted with the exact same Scrabble board and given the exact same letters. The beginners give a wide range of answers so they score high on originality. These answers are worth varying numbers of points in the game so they get an intermediate score for appropriateness. The experts, on the other hand, all give the one-and-only seven-letter word, so according to the rules of Scrabble they all get 50 extra points; thus their answers uniformly exhibit a high degree of appropriateness. However, because they all converge on the same answer, they collectively score zero for originality. Since according to the standard definition, creativity requires both appropriateness and originality, the beginners are rated as more creative than the experts.

The tension between creativity and expertise dates back to the writings of William James (1890, p. 456) and is still in evidence (Frensch & Sternberg, 1989; Weisberg, 2006). A natural consequence of this tension is to emphasize the originality component of creativity, but surely something is wrong with our conception of creativity if it does not definitively lead us to rate a brilliant, high-scoring Scrabble play as more creative than one that is mediocre but original. Clearly the phenomenon is not peculiar to Scrabble. An analogous situation undoubtedly exists for Chess, and other creative endeavors. The phenomenon not only highlights where our current definition of creativity falls short, but leads straightforwardly to a proposal for how it can be improved.

## 2. Applying Adaptive Landscapes to Creativity

It is proposed that whether or not the 'original and appropriate' definition of creativity holds depends on the *adaptive landscape* associated with the creative task. Adaptive landscapes, also known as fitness landscapes, were introduced by Sewell Wright (1932) as a way of depicting the relationship between genotype and reproductive success. Adaptive landscapes have proven useful for understanding biological phenomena such as epistasis and pleiotropy (Kauffman, 1993), and the notion has been generalized and applied to fields outside of biology such as combinatorial optimization problems (e.g. Mitchell, 1996), and to the evolution of creative ideas through culture (e.g. Gabora, 1995, 1996). They provide a straightforward way of clarifying the relationship between creativity, originality, appropriateness, and the nature of the task.

Consider a graph with different possible ideas plotted along the *x* and *y* axes, and a distance metric such that the closer together they are, the more similar they are. In general, creative ideas differ along many dimensions, but for simplicity we assume there are two. A measure of the idea's appropriateness is represented on the *z* axis. The set of all possible ideas, their degree of similarity, and a measure of their relative appropriateness, constitute an *adaptive landscape*. Thus an adaptive landscape differs from the notion of a state space or a conceptual space in associating with each possibility an assessment of it.

First we consider the situation that the adaptive landscape associated with a problem or task is flat, such that all possibilities are equally appropriate, as depicted in Figure 1a. To return to the Scrabble example, this would be the case if, given the state of the game and one's set of letters, all possible plays give the same number of points. Finding any solution to a flat-landscape problem, i.e. any point on this landscape, is more creative than finding no solution at all. Therefore, all solutions are at least somewhat creative. Those that are chosen infrequently are more original, and thus more creative, than those chosen frequently. Since in this situation it is not possible to increase creativity by increasing appropriateness, the standard definition of creativity does not hold. For flat adaptive landscape tasks, the only way to be more creative is to be more original.

[PLACE FIGURE 1 ABOUT HERE]

Now let us consider the situation in which the adaptive landscape associated with a creative task is single-peaked, with one solution that is clearly superior to the rest, and each step closer to the optimal solution is higher in appropriateness. This is depicted in Figure 1b. The fewer and more pronounced the peaks in the adaptive landscape associated with a creative task, the greater the extent to which originality is inversely correlated with appropriateness. As one converges on more appropriate solutions, they become less original, and indeed such tasks are often said to require *convergent thinking*. In the extreme, consider a task with 'needle in the haystack' adaptive landscape, for which one idea is superior to the rest, and every other idea, no matter how similar it is to the superior idea, is equally poor. This is depicted in Figure 1c. Here the tradeoff between originality and appropriateness is even more extreme. It is more or less exemplified by the Scrabble example given at the beginning of the article, wherein expert players converge on a single optimal solution and beginners give a variety of vastly inferior solutions. Since for problems with single-peaked adaptive landscapes it is not possible to increase appropriateness without compromising originality and vice versa, once again the standard definition of creativity does not hold. Some standard creativity tests, such as the Remote Associates Test (Mednick, 1962), and many tests of analogical problem solving (e.g. Gick & Holyoak, 1980), as well as most intelligence tests, involve problems of this sort.

Last we consider the situation that the adaptive landscape associated with the problem is rugged, with multiple peaks of comparable height separated by valleys, as depicted in Figure 1d. The ruggedness of the adaptive landscape with respect to Scrabble is increased by the fact that particular squares of the board allow one to double or triple the points conferred by a letter or word. The result is that similar plays may give quite different numbers of points. When the adaptive landscape associated with a problem is rugged, it is possible to increase the creativity of an idea by increasing either its originality or its appropriateness, so the standard definition of creativity does hold. Indeed, most standard tests of creativity, particularly tests of *divergent thinking* (e.g. Getzels & Jackson, 1962; Guilford, 1967; Torrance, 1962, 1974; Wallach & Kogan, 1965), have rugged adaptive landscapes because there are many acceptable answers though some answers would be inferior or unacceptable. In these tests creativity is often rated in terms of both originality and *fluency,* i.e. the number of different acceptable answers one gives. Thus for example, if one is asked to think of as many uses as possible for a brick, one's fluency score depends on how many uses for a brick one comes up with.

This analysis shows that the standard definition of creativity is not applicable to all creative tasks. The flatter the adaptive landscape associated with a task, the greater the extent to which it makes sense to rate creativity in terms of originality. The more single-peaked the adaptive landscape associated with the task, the greater the extent to which it makes sense to rate creativity in terms of appropriateness. For tasks with rugged adaptive landscapes, the greater the extent to which it makes sense to rate creativity in terms of both originality and appropriateness.

## 3. Surprise and the 'Aha' Experience from an Adaptive Landscapes Perspective

We saw that Boden (2004) adds surprise as a third criterion for creativity. It has been suggested that the 'Aha' experience associated with insight involves the formation of new concept combinations (Koestler, 1967; Mednick, 1962). Thagard and Stewart (2011) propose that surprise results when the *convolution* or 'twisting together' of mental representations co-occurs with a second convolution event involving an appraisal and a physiological response. It has been proposed that unusual concept combinations arise through the firing of neural cell assemblies that would not normally participate in the activation of a given concept, but that do in an unusual context, when one is in a divergent mode of thought (Gabora, 2010). The adaptive landscapes perspective suggests that the degree of surprise is correlated with the slope of the adaptive landscape at the transition point from the conception of the task that does not incorporate the concept combination to a subsequent conception of the task that does. It seems reasonable that the feeling of not just surprise but exhilaration associated with insight into a particularly challenging task has to do with intuiting that the new concept combination is in the vicinity of, though not yet at the top of, a substantial peak.

## 4. Redefining the Relationship between Creativity and Expertise

Let us return to the tension between creativity and expertise. Expertise undoubtedly aids creativity. Experts are better than beginners at detecting and remembering domain-relevant patterns, and are more adept at generating effective problem representations (Ericsson & Charness, 1994) and, when necessary, revising initial hypotheses (Alberdi, Sleeman & Korpi, 2000); indeed it is often said that it takes approximately a decade to master a creative domain (Hayes, 1989). Yet entrenchment in established perspectives and approaches may make experts more prone than beginners to set, functional fixedness, and confirmation bias (Frensch & Sternberg, 1989; Mumford et al., 2004).[2]

The adaptive landscape perspective helps tease apart beneficial and detrimental sources of unoriginality associated with expertise. A beneficial lack of originality is unavoidable when the adaptive landscape is single-peaked, or has few peaks, as in the above example. For example, this is the situation in the Scrabble play in which beginners give a variety of original answers while experts converge on the same seven-letter-word answer. In such cases, although experts converge on the same answer, they can legitimately be said to exhibit a greater degree of creativity than the beginners. The tendency toward unoriginality here reflects constraints on creativity due to the shape of the adaptive landscape, to which the expert is better able to respond. This is expected to be the case often with respect to domains for which people acquire expertise, because if the adaptive landscape with respect to a particular task domain is essentially flat, there is little incentive to acquire expertise in that domain. The beneficial effects of expertise may be accentuated if the adaptive landscape is static, or changes in a manner that is, for the expert, predictable.

The second source of unoriginality associated with expertise is lack of originality due to susceptibility to phenomena such as set, functional fixedness, and confirmation bias. This may be more likely when the adaptive landscape associated with the task is rugged (with many peaks and valleys) because there is greater risk of getting stuck on a local maximum and never finding the global maximum. In this case, the tendency toward unoriginality does not reflect constraints arising due to the shape of the adaptive landscape, for multiple appropriate, original answers are possible. Lack of originality here reflects limitations on the part of the problem solver rather than constraints imposed by the problem itself. Here the beginner, approaching the task from a fresh perspective, may succeed where the expert fails, particularly if the adaptive landscape is susceptible to chaotic and unpredictable change.

## 5. Characterizing Task Domains on the Basis of Adaptive Landscapes

Adaptive landscapes offer a conceptual framework for characterizing problems, tasks, and games. One parameter that distinguishes them is whether different individuals who engage in the task, or players in the game, start off with the same adaptive landscape. This is the case, for example, in Chess, but not in Scrabble, where each player begins with different sets of letters, which afford different possible plays.

A second parameter is the extent to which one individual's moves affect the topology of the adaptive landscape for other individuals. This is the case for board games and athletic games; indeed one could argue that this is what *makes* a game a game, rather than some other sort of task. It is less the case for other problem solving or creative tasks, although in those too, individuals affect the topology of the adaptive landscape of the task for other individuals to the extent that they inspire, compete with, help, or hinder one another.

A third parameter is the extent to which different individuals' adaptive landscapes diverge over the duration of the task or game. In Scrabble, although players begin with different adaptive landscapes, as discussed above, their adaptive landscapes do not progressively differentiate; indeed as letters get used, and the structure of the board builds up, different players' adaptive landscapes may become constrained in similar ways. In other creative tasks, it may be the case that as one embarks further in a particular direction, each move or action changes the utility of possible subsequent moves or actions, with the result that different individuals' adaptive landscapes diverge over the course of the task or game.

## 6. Redefining the Relationship between Creativity in Science and the Arts

The notion of adaptive landscapes provides a tool for conceptualizing the distinction between artistic and scientific creativity. It seems self-evident that science is more constrained than art, and that two scientists are more likely to converge on the same result than two artists are to produce the same painting, or two writers are to produce the same novel. Artistic pursuits are often referred to as ill-defined or open-ended, and are said to "admit multiple "good enough" solutions, rather than one "correct" answer" (Kozbelt, Beghetto, & Runco, 2011, p. 33). One might be tempted to think that the adaptive landscapes associated with scientific advances are less rugged than those associated with artistic works. Another possibility is that the adaptive landscape that guides progress in science is the roughly same for each individual, while the adaptive landscape that guides the generation of works of art is different for each individual. If the goal of science is to understand something about the world that was not known before, since we all live in the same physical world, what constitutes an appropriate scientific work will be more or less the same for each individual. But if the goal of a work of art is to identify, express, and thereby externalize what one has experienced, then since we all have different experiences, and different feelings and responses to these experiences, what constitutes an appropriate work of art differs for each individual. If this proposal is correct, then although it may seem that the adaptive landscapes for artistic projects are more rugged than that for scientific projects, it is merely the case that the peaks and valleys are aligned for different individuals with respect to scientific projects, but not with respect to artistic projects. In other words, artistic projects are no more open-ended than science projects; for a given artist working on a given project, the peaks in the adaptive landscape may be as pronounced as they are for a scientist. However, because the adaptive landscape reflects unique aspects of the artists' experience and personality, the location of these peaks is more idiosyncratic, and less transparent to others.

This is consistent with findings that talented artists, like scientists, have a clear sense of what constitutes a successful finished product (Feinstein, 2006). It is also consistent with evidence that the advantages associated with expertise mentioned previously—such as enhanced ability to remember domain-relevant patterns and to generate effective problem representations—can facilitate performance even in creative domains such as music composition and art (Ericsson, 1999; Kozbelt, 2008).

## 7. Conclusions

A one-size-fits-all definition of creativity is not possible. Since divergent thinking tasks tend to have rugged adaptive landscapes, it is reasonable to define and assess creativity in terms of both originality and appropriateness. The flatter the adaptive landscape, the more reasonable it is to consider originality and fluency only. For convergent thinking tasks, with single-peaked landscapes, it is reasonable to consider appropriateness only. The view that creativity requires both appropriateness and originality may have led to an over-reliance on divergent thinking tests with less than stellar predictive validity (Baer, 1993; Cooper, 1991; Kogan & Pankove, 1974), perhaps because they require little in the way of in-depth contemplation.

Divergent thought may be used not just in the generation of multiple off-the-cuff solutions, such as takes place in most divergent thinking tests, but in the generation and development of a single complex idea, of the sort that is measured by the Consensual Assessment Technique (Amabile, 1982). There is support for the hypothesis that complex creative ideas arise in a state of potentiality (i.e., 'half-baked'), that require dedicated effort to work out (Gabora & Saab, 2011). This suggests that what differentiates the sorts of tasks used in typical divergent thinking tests from more complex creative tasks is that the first involves forming combinations that constitute full-fledged solutions from the start, so they land creator at the top of one of many modest peaks, whereas the second involves forming incomplete combinations that land the creator in the vicinity of a pronounced peak. What distinguishes complex creative tasks from other complex problem solving tasks may be not the shape of the adaptive landscape but the frequency which the concept combinations necessary to come up with possibilities in this landscape are carried out, i.e. how well-trodden it is.

An intuitive appreciation of the distinctive adaptive landscapes associated with different problems is implicit in the different measures employed in different tests of creativity. It would be possible to systematically tailor the assessment of creativity to the constraints and affordances of the task by weighting different measures—such as accuracy, originality, appropriateness, and fluency—in a way that matches the adaptive landscape. One could then select batteries of tests for which the adaptive landscapes have different topologies so as to obtain a refined and well-rounded assessment of creativity.

It is sometimes said that science involves discovering something that already existed whereas art involves bringing something into existence. The adaptive landscapes perspective suggests an alternative: the scientist is navigating a space of possibilities that is more or less the same for everyone, while the artist is navigating a space of possibilities that, though equally real, is at least to some degree inaccessible to others.

## References

Alberdi, E., Sleeman, D., & Korpi, M. (2000). Accommodating surprise and taxonomic tasks: The role of expertise. *Cognitive Science*, *24*(1), 53-91.

Albert, R. S., & Runco, M. A. (1989). Independence and cognitive ability in gifted and exceptionally gifted boys. *Journal of Youth and Adolescence*, 18, 221–230.

Amabile, T. M. (1982). Social psychology of creativity: A consensual assessment technique. *Journal of Personality and Social Psychology*, *43,* 997–1013.

Amabile, T. (1996). *Creativity in context*. Boulder, CO: Westview Press.

Baer, J. (1993). Why you shouldn't trust creativity tests. *Educational Leadership*, *51*(4), 80–83.

Boden, M. (2004). *The creative mind: Myths and mechanisms* (2nd ed.). London: Routledge.

Campbell, D. T. (1960). Blind variation and selective retention in creative thought as in other knowledge processes. *Psychological Review*, 67, 380–400.

Cooper, E. (1991). A critique of six measures for assessing creativity. *Journal of Creative Behavior*, *25*, 194–204.

Dane, E. (2010). Reconsidering the trade-off between expertise and flexibility: A cognitive entrenchment perspective. *The Academy of Management Review, 35*(4), 579-603.

Ericsson, K. A. (1999). Creative expertise as superior reproducible performance: Innovative and flexible aspects of expert performance. *Psychological Inquiry, 10,* 329–333.

Feinstein, J. S. (2006). *The nature of creative development*. Stanford: Stanford University Press.

Feist, G. (2011). The function of personality in creativity. In (J. Kaufman & R. Sternberg, Eds.) *Cambridge Handbook of Creativity*. (pp. 113-130). Cambridge UK: Cambridge University Press.

Feldman, D. H., Csikszentmihalyi, M., & Gardner, H. (1994). *Changing the world: A framework*

*for the study of creativity*. Westport, CT: Praeger Publishers.

Frensch, P. A., & Sternberg, R. J. (1989). Expertise and intelligent thinking: When is it worse to know better? In R. J. Sternberg (Ed.) *Advances in the psychology of human intelligence* (Vol. 5). Hillsdale, NJ: Erlbaum.

Gabora, L. (1995). Meme and variations: A computer model of cultural evolution. In (L. Nadel & D. Stein, Eds.) *1993 Lectures in Complex Systems,* Addison-Wesley, 471-486.

Gabora, L. (1996). A day in the life of a meme. *Philosophica, 57*, 901-938.

Gabora, L. (2010). Revenge of the 'neurds': Characterizing creative thought in terms of the structure and dynamics of human memory. *Creativity Research Journal, 22*(1), 1-13.

Gabora, L. & Saab, A. (2011). Creative interference and states of potentiality in analogy problem solving. *Proceedings of the Annual Meeting of the Cognitive Science Society*. July 20-23, 2011, Boston MA.

Getzels, J. W., & Jackson, P. W. (1962). *Creativity and intelligence: Explorations with gifted students*. New York: Wiley.

Gick, M. L., & Holyoak, K. J. (1980). Analogical problem solving. *Cognitive Psychology*, 12, 306–355.

Hayes. J. R. (1989). Cognitive processes in creativity. In J. A. Glover. R. R. Ronning. & C. R. Reynolds (Eds.). *Handbook of creativity* (pp. 135-145). New York: Plenum Press.

James, W. (1890). Great men, great thoughts, and the environment. *Atlantic Monthly, 46,* 441-459.

Kauffman, S. (1993). Origins of order. Oxford: Oxford University Press.

Kaufman, J. C., & Sternberg, R. J. (2007). Resource review: Creativity. *Change*, 39, 55–58.

Koestler, A. (1967). *The act of creation*. New York: Dell.

Kogan, N., & Pankove, E. (1974). Long-term predictive validity of divergent-thinking tests: Some negative evidence. *Journal of Educational Psychology*, 66, 802–810.

Kozbelt, A. (2008). Longitudinal hit ratios of classical composers: Reconciling “Darwinian” and expertise acquisition perspectives on lifespan creativity. *Psychology of Aesthetics, Creativity, and the Arts, 2*, 221–235.

MacKinnon, D. W. (1975). IPAR’s contribution to the conceptualization and study of creativity. In I. A. Taylor & J. W. Getzels (Eds.), *Perspectives in creativity*. Chicago: Adaline.

Mednick, S. A. (1962). The associative basis of the creative process. *Psychological Review*, 69, 220–232.

Mitchell, M. (1996). *An introduction to genetic algorithms*. Cambridge, MA: MIT Press.

Moran, S. (2011). The roles of creativity in society. In (J. Kaufman & R. Sternberg, Eds.) *Cambridge Handbook of Creativity*. (pp. 74-90). Cambridge UK: Cambridge University Press.

Mumford, M. D., Blair, C., Dailey, L., Leritz, L. E., & Osborn, H. K. (2004). Errors in creative thought: Cognitive biases in a complex processing activity. *Journal of Creative Behavior, 40,* 75-109.

Mumford, M. D., & Gustafson, S. G. (1988). Creativity syndrome: Integration, application, and innovation. *Psychological Bulletin, 103,* 27–43.

Runco, M. A. (2004). Creativity. *Annual Review of Psychology, 55*, 657–687.

Sternberg, R. J. (1988). A three-faceted model of creativity. In R. J. Sternberg (Ed.), *The nature of creativity* (pp. 125–147). Cambridge, England: Cambridge University Press.

Thagard, P. & Stewart, T. C. (2011). The AHA! Experience: Creativity through emergent binding in neural networks. *Cognitive Science, 35,* 1-33.

Torrance, E. P. (1962). *Guiding creative talent*. Englewood Cliffs, NJ: Prentice-Hall.

Torrance, E. P. (1974). *Torrance Tests of Creative Thinking: Norms-technical manual*. Bensenville, IL: Scholastic Testing Service.

Wallach, M. A., & Kogan, N. (1965). *Modes of thinking in young children: A study of the creativity-intelligence distinction*. New York: Holt, Rinehart and Winston.

Weisberg, R. (2006). *Creativity: Understanding innovation in problem solving, science, invention, and the arts*. Hoboken NJ: John Wiley & Sons.

Wright, S. (1932). The roles of mutation, inbreeding, crossbreeding, and selection in evolution. In *Proceedings of the Sixth International Congress on Genetics*, pp. 355–366.

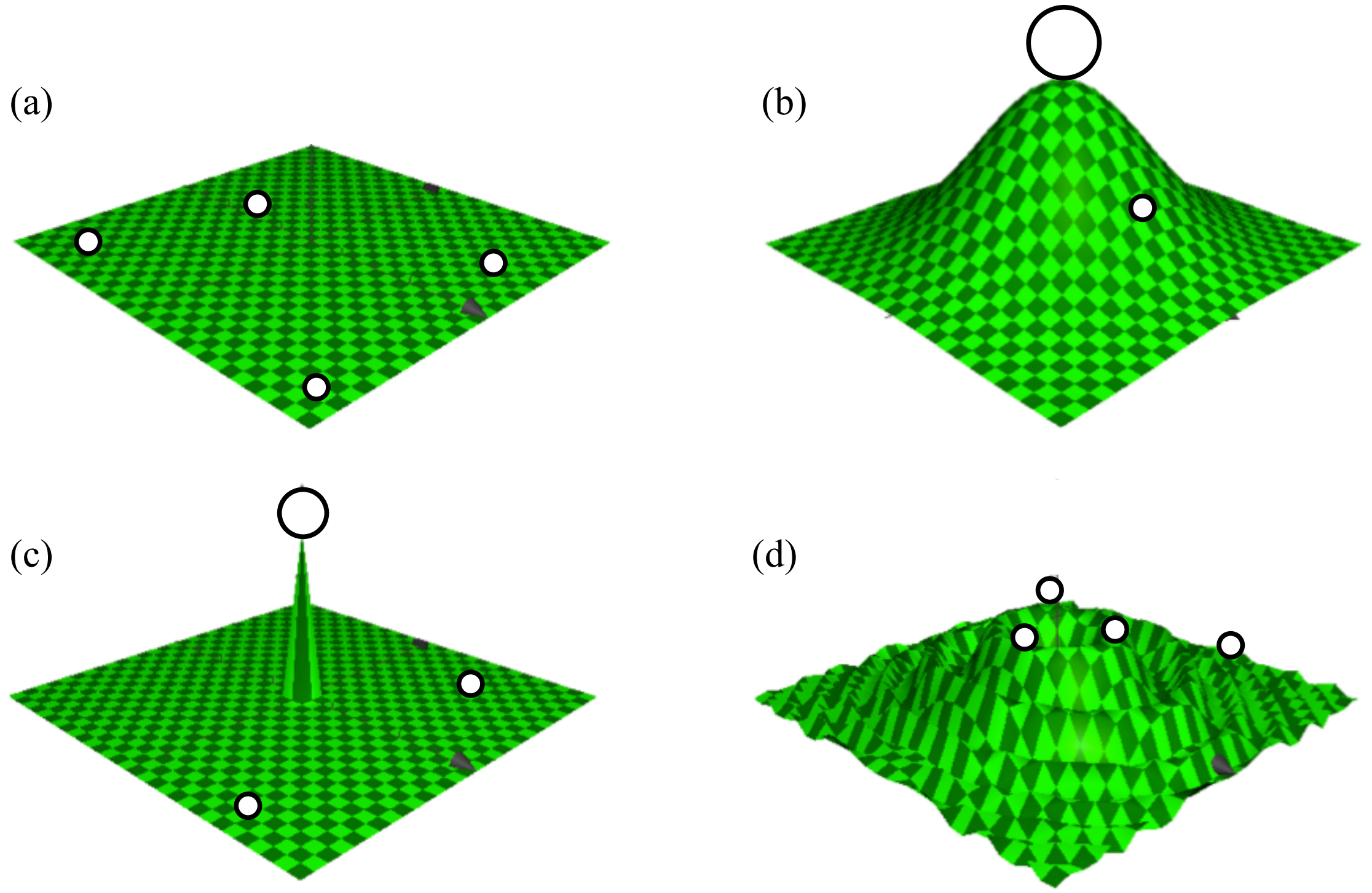

Figure 1. Schematic illustration of the adaptive landscapes associated with different tasks. Size of white circle represents the number of individuals who give a particular solution (thus it is inversely correlated with originality). (a) For a task with a *flat adaptive landscape*, all solutions are rated equally, so creativity is a function of originality only. (b) For a task with a *single-peaked adaptive landscape*, there is one optimal solution, and the more similar a solution is to that optimal solution, the more creative it is. Creativity is primarily a function of appropriateness. (c) For a task with a *needle in the haystack adaptive landscape*, there is one optimal solution, and all others solutions are equally inferior. Creativity is a function of appropriateness only. (d) For a task with a *rugged adaptive landscape*, with many peaks and valleys, creativity is a complex function of both originality and appropriateness.

---

[1] I became aware of the Scrabble manifestation of this phenomenon during a conversation with one of the 100 top-rated Scrabble players in the world (Jesse Matthews; Pers. Comm., 2010).

[2] There are, however, methods for circumventing some of the costs associated with expertise (e.g. Dane, 2010).